\documentclass[
aip,
jmp,
numerical,
 amsmath,amssymb,
 reprint,%
]{revtex4-1}
\usepackage{graphicx}
\usepackage{dcolumn}
\usepackage{bm}
\usepackage{amsmath}
\usepackage{float}
\usepackage{subcaption}
\usepackage{makecell}
\usepackage{comment}
\usepackage{caption}
\renewcommand{\thesection}{\arabic{section}} 
\renewcommand{\thesubsection}{\thesection.\arabic{subsection}}

\usepackage[breaklinks=true,colorlinks=true,linkcolor=blue,urlcolor=blue,citecolor=blue]{hyperref}
\numberwithin{equation}{section}
\renewcommand{\theequation}{\arabic{section}.\arabic{equation}}
\begin{document}
\title{Algebraic derivation of the Energy Eigenvalues for the quantum oscillator defined on the Sphere and the Hyperbolic plane}
\author{Atulit Srivastava}
 \email{srivastava.atulit@gmail.com}
\affiliation{ 
 Department of Physics and Astronomy, KU Leuven, Belgium
}
\author{S.K. Soni}
 \affiliation{S.G.T.B. Khalsa college, University of Delhi, India.}
\date{\today}
\begin{abstract}
We give an algebraic derivation of the eigenvalues of energy of a quantum harmonic oscillator on the surface of constant curvature, i.e. on the sphere or on the hyperbolic plane. We use the method proposed by Daskaloyannis for fixing the energy eigenvalues of two-dimensional (2D) quadratically superintegrable systems by assuming that they are determined by the existence of finite-dimensional representation of the polynomial algebra of the motion integral operators. The tool for realizing representations is the deformed parafermionic oscillator. The eigenvalues of energy are calculated and the result derived by us algebraically agrees with the known energy eigenvalues calculated by classical analytical methods. This assertion which is the main result of this article is demonstrated by a detailed presentation. We also discuss the qualitative difference of the energy spectra on the sphere and on the hyperbolic plane.  
\end{abstract}
\maketitle
\section{Introduction}
We discuss a quadratic algebraic approach to the classical\cite{carinena2004non} and the quantum\cite{article,ranada2014quantum,carinena2007quantum,carinena2007quantum2,quesne2015update} harmonic oscillators on the sphere and the hyperbolic plane leading ultimately to their spectral properties without invoking the Schrodinger equation; the energy eigenvalues were calculated by using classical means in references [\onlinecite{carinena2007quantum}] and [\onlinecite{carinena2007quantum2}, \onlinecite{quesne2015update}] wherein special function solutions of a spectral problem appear. These nonlinear classical and quantum models extend to two spatial dimensions a well known classical nonlinear oscillator \cite{mathews1974unique,lakshmanan2012nonlinear} with an exactly solvable quantum analog \cite{schulze2012special,carinena2004one,carinena2007quantum3}.
The non-linearity parameter $\lambda$ which enters the definition of both a nonlinear potential and a position dependent mass in the model of references [\onlinecite{mathews1974unique}] and [\onlinecite{lakshmanan2012nonlinear}] can be interpreted in the 2D model of Cari{\~n}ena et al \cite{carinena2004non} as the negative of the constant Gaussian curvature $\kappa$ of the underlying two-dimensional space (i.e. $\lambda = -\kappa$), thereby showing that their model describes a harmonic oscillator on the sphere for negative $\lambda$ and on the hyperbolic plane for positive $\lambda$.

Cari{\~n}ena et al\cite{carinena2004non} showed that their 2D system is superintegrable and that its Hamilton-Jacobi equation is separable in three different coordinate system in agreement with the fact that superintegrable systems are Hamiltonian systems with more integrals of motion than the degrees of freedom. In two dimensions, the maximum number of functionally independent integrals is three, and that is the case in reference [\onlinecite{carinena2004non}] we are considering. In references [\onlinecite{carinena2007quantum},\onlinecite{carinena2007quantum2}] the energy eigenvalues and wave-function for the quantum analog of this classical model were determined by solving the spectral problem in the corresponding three coordinate systems where the Schrodinger equation becomes separable.

In the present work we shed new light on the 2D classical and quantum models of the harmonic oscillator by following a quantum algebraic approach due to Daskaloyannis\cite{daskaloyannis2001quadratic} to calculate energy eigenvalues of the 2D quadratically superintegrable systems. The method of reference [\onlinecite{daskaloyannis2001quadratic}] fixes the energy eigenvalues by assuming that they can be determined by the existence of a finite-dimensional representation of the polynomial algebra of the motion integral operators. The entire calculation proceeds by using the tool of deformed parafermionic oscillator algebra \cite{quesne1994generalized}. The energy eigenvalues can be found by solving the appropriate algebraic equations with two unknown parameters $u$ and $E$ to be determined. The energy eigenvalues were calculated by the authors of references [\onlinecite{carinena2007quantum}] and [\onlinecite{carinena2007quantum2}, \onlinecite{quesne2015update}]. In this article we calculate the eigenvalues of energy and find that these eigenvalues of energy are the same as the eigenvalues which were calculated by classical analytical techniques.
It is worth noting a preliminary attempt to calculate the energy spectrum using the analogy to the harmonic oscillator on the Euclidean plane in the reference  [\onlinecite{bonatsos1994deformed}].

The structure of the paper is as follows. In Section \ref{Quadratic Poisson Algebra}, we consider a Hamiltonian system which is more general than the curved space generalization of the classical harmonic oscillator and review the general form of the quadratic Poisson algebra presented in reference [\onlinecite{daskaloyannis2001quadratic}] for the quadratically superintegrable system. In Section \ref{Quadratic Poisson algebra for the 2D-nonlinear harmonic oscillator}, we then apply these general considerations to arrive at the special form of the quadratic Poisson algebra for the 2D nonlinear harmonic oscillator. In Section \ref{Quadratic Associative algebra and its deformed oscillator realization}, we give the quantum analogue of the quadratic Poisson algebra considered in Section \ref{Quadratic Poisson Algebra}, the corresponding quadratic associative algebra. We then express the Casimir operator for this algebra in terms of the Hamiltonian and give the realization of the quadratic algebra in terms of the deformed parafermionic oscillator algebra. The finite dimensional representations of the quadratic algebra are generated by using the technique of the deformed parafermionic algebra. Finally, the problem is reduced to the solution of a system of two algebraic equations. In Section \ref{Quadratic associative algebra and the energy eigenvalues for the 2D quantum harmonic oscillator}, we give the quadratic associative algebra and the energy eigenvalues for the 2D quantum harmonic oscillator determined by solving the appropriate algebraic equations. In Section \ref{Discussion and Conclusion}, we give a precise and a detailed discussion of the comparison with the results of the analytical calculations. Our main conclusions are given at the end of that section.

\section{Quadratic Poisson Algebra}\label{Quadratic Poisson Algebra}
In this section we consider a general superintegrable system in a two-dimensional space(not necessarily Euclidean) with a scalar potential. We will assume that we have a quadratic Hamiltonian and due to superintegrability the system possesses two more integrals of motion A and B that are quadratic functions of canonical momenta too. Since A and B are constants of motion they satisfy the following Poisson bracket relations,
\begin{align}
\{H,A\}=\{H,B\}=0.\label{eq:2.1}
\end{align}
Since H, A, B are quadratic in momenta, we can expect them to generate a quadratic algebra like that in reference [\onlinecite{daskaloyannis2001quadratic}]. We set 
\begin{align}
\{A,B\}&=C,  \nonumber  \\
\{A,C\}&=\alpha A^2+ 2\gamma AB + \delta A + \epsilon B + \zeta,  \nonumber \\
\{B,C\}&= a A^2+\rho B^2+2\sigma AB+dA+\eta B+z.\label{eq:2.2}
\end{align}
Since we have $\{C,\{A,B\}\}=0$ the Jacobi identity reduces to
\begin{align}
\{A,\{B,C\}\}=\{B,\{A,C\}\}.\label{eq:2.3}    
\end{align}
Thus the Jacobi identity implies $\rho=-\gamma$, $\sigma=-\alpha$, $\eta=-\delta$ and we obtain the quadratic algebra
\begin{align}
\{A,B\}&=C, \nonumber \\
\{A,C\}&=\alpha A^2+ 2\gamma AB + \delta A + \epsilon B + \zeta,  \nonumber \\
\{B,C\}&= a A^2-\gamma B^2-2\alpha AB+dA-\delta B+z.\label{eq:2.4}
\end{align}
The coefficients $a$, $\alpha$ and $\gamma$ are constants, but the other ones can be polynomials in the Hamiltonian H. The degrees of these polynomials are dictated by the fact that $H, A$ and $B$ are second order polynomials in the momenta. Hence, C can be a third order polynomial. We have that $\delta$, $\epsilon$ and $d$ are each equal to at most a linear function of H whereas  $\zeta$ and $z$ are each equal to at most a quadratic function of H. A Casimir operator $K$ of a polynomial algebra is defined as a Poisson operator commuting with all the elements of the algebra. For the algebra (\ref{eq:2.4}) this means that
\begin{align}
\{K,A\}=\{K,B\}=\{K,C\}=0,    \label{eq:2.5}
\end{align}
and this implies 
\begin{align}
K=C^2 - 2\alpha A^2B-2\gamma AB^2-2\delta AB- \epsilon B^2-2\zeta B+\frac{2}{3}a A^3 +dA^2 +2zA.\label{eq:2.6}
\end{align}
Thus K is a polynomial of degree 6 in the momenta. Since the Hamiltonian H also satisfies relations (\ref{eq:2.5}) we can expect K to be a polynomial in H and we write
\begin{align}
K = k_0+k_1 H+k_2 H^2 +k_3 H^3,  \label{eq:2.7}  
\end{align}
where $k_0$, $k_1$, $k_2$ and $k_3$ are constants. Note that (\ref{eq:2.7}) together with (\ref{eq:2.6}) represents a polynomial relation between the integrals $H$, $A$, $B$ and $C$ in agreement with the fact that the integral of motion $C$ is not independent from the integrals $H$, $A$ and $B$. Therefore the integrals of motion of a quadratically superintegrable two dimensional system close to form the quadratic Poisson algebra (\ref{eq:2.4}), corresponding to a Casimir equal to at most a cubic function of the Hamiltonian (\ref{eq:2.7}).  

\section{Quadratic Poisson algebra for the 2D nonlinear harmonic oscillator}\label{Quadratic Poisson algebra for the 2D-nonlinear harmonic oscillator}
Now let us consider the superintegrable Hamiltonian \cite{carinena2004non} for the curved space generalization of the harmonic oscillator on the 2D sphere and the hyperbolic plane
\begin{align}
H=  \frac{1}{2}\left[p_x^2+p_y^2-\kappa(x p_x+y p_y )^2 \right]+\frac{\omega^2}{2} \left(\frac{x^2+y^2}{1-\kappa(x^2+y^2)}\right). \label{eq:3.1}   
\end{align}
The Hamiltonian can also be written as 
\begin{align}
H=H_1+ H_2+\kappa H_3,    \label{eq:3.2}
\end{align}
where $H_1$, $H_2$ and $H_3$ are the three independent integrals of motion that are given by
\begin{align}
H_1&=\frac{1}{2}\left(\left[1-\kappa (x^2+y^2) \right] p_x^2+\omega^2 \frac{x^2}{1-\kappa(x^2+y^2)}\right)\label{eq:3.3},\\
H_2&=\frac{1}{2}\left(\left[1-\kappa (x^2+y^2) \right] p_y^2+\omega^2 \frac{y^2}{1-\kappa(x^2+y^2)}\right)\label{eq:3.4},\\
H_3&=\frac{1}{2} (x p_y-y p_x )^2.\label{eq:3.5}
\end{align}
 Taking A as $2 H_1$ and B as $2 H_2$ we can identify the classical coefficients which characterize the corresponding algebra (\ref{eq:2.4}). They are given by
\begin{align}
\alpha&=-8,\nonumber\\
\gamma&=-8\kappa,\nonumber\\
\delta&=16H,\nonumber\\
\epsilon&=-16\omega^2,\nonumber\\
\zeta&=a=d=z=0.\label{eq:3.6}
\end{align}
The value of the Casimir calculated from (\ref{eq:2.6}) is
\begin{align}
K=0.    \label{eq:3.7}
\end{align}
It is essential to study the quadratic Poisson algebra and its Casimir as we will observe in the next section that it correspond to the lowest order terms in $\hslash$ of the quadratic associative algebra and its Casimir operator in the corresponding quantum system.

It is worth remarking that the above algebra is between $A$, $B$ and $C$ is the direct consequence of the underlying Poisson algebra between integrals $C_i$ defined as follows 
\begin{align}
 C_1&=H, \\
 C_2&=H_1=\frac{1}{2}\left([1-\kappa (x^2+y^2) ] p_x^2+\omega^2 \frac{x^2}{1-\kappa(x^2+y^2)}\right),\\ 
 C3&=H_2=\frac{1}{2}\left([1-\kappa (x^2+y^2) ] p_y^2+\omega^2 \frac{y^2}{1-\kappa(x^2+y^2)}\right),\\
 C_4&= (x p_y-y p_x ),\\
 C_5&=\left([1-\kappa (x^2+y^2) ] p_xp_y+\omega^2 \frac{xy}{1-\kappa(x^2+y^2)}\right).
\end{align}
So, we have the following two constraint functionals
\begin{align}
F_1&=C_1-\left(C_2+C_3+\frac{\kappa}{2}C_4^2\right),   \\
F_2&=2C_2C_3-\frac{\alpha^2}{2}C_4^2-\frac{1}{2}C_5^2,
\end{align}
such that they vanish on the space of solutions.
Following Tegman\cite{teugmen2006nambu}, the non-vanishing Poisson brackets between the $C_i$ can be written in terms of the constrained functionals
\begin{align}
\{C_2, C_4\} &= -\frac{\partial(F_1,F_2)}{\partial(C_3,C_5)}=-C_5,\\
\{C_2, C_3\} &= \frac{\partial(F_1,F_2)}{\partial(C_4,C_5)}=\kappa C_4C_5,\\
\{C_2, C_5\} &= \frac{\partial(F_1,F_2)}{\partial(C_3,C_4)}=\alpha^2C_4+2\kappa C_2C_4,\\
\{C_3, C_4\} &= \frac{\partial(F_1,F_2)}{\partial(C_2,C_5)}=C_5,\\
\{C_3, C_5\} &= -\frac{\partial(F_1,F_2)}{\partial(C_2,C_4)}=-\alpha^2C_4-2\kappa C_3C_4,\\
\{C_4, C_5\} &= \frac{\partial(F_1,F_2)}{\partial(C_2,C_3)}=(2C_3-2C_2).
\end{align}

\section{Quadratic Associative algebra and its deformed oscillator realization }\label{Quadratic Associative algebra and its deformed oscillator realization}
In this section, we review the quantum version of the quadratic Poisson algebra which is the quadratic associative algebra of operators for the quantum 2D quadratically superintegrable systems. The integrals A and B are now independent motion integral operators. The quadratic associative algebra for the generators \{A, B, C\} generated by motion integral operators of the 2D quadratically superintegrable systems satisfy the following commutation relations\cite{daskaloyannis2001quadratic}
\begin{align}
[A,B]&=C\label{eq:4.1},\\
[A,C]&= \alpha A^2+\gamma\{A,B\}+\delta A+\epsilon B+\zeta \label{eq:4.2},\\    
[B,C]&= a A^2-\gamma B^2-\alpha\{A,B\}+dA-\delta B+z. \label{eq:4.3}
\end{align}
The Casimir operator for this algebra is given by
\begin{align}
K&=C^2-\alpha\{A^2,B\}-\gamma\{A,B^2 \}\nonumber \\
&+(\alpha\gamma-\delta)\{A,B\}+(\gamma^2-\epsilon) B^2+(\gamma\delta-2\zeta)B\nonumber\\
&+\frac{2}{3} aA^3+(d+\frac{a\gamma}{3}+\alpha^2) A^2+(\frac{a\epsilon}{3}+\alpha\delta+2z)A.\label{eq:4.4}
\end{align}
The quadratic algebra Q(3) (\ref{eq:4.1}), (\ref{eq:4.2}) and (\ref{eq:4.3}) is realized in terms of the deformed oscillator algebra\cite{daskaloyannis2001quadratic} $\{\mathcal{N}, b^\dagger, b\}$ satisfying the following equations:
\begin{align}
[\mathcal{N},b^{\dagger}]&=b^{\dagger}\nonumber,\\
[\mathcal{N},b]&=-b\nonumber,\\
b^{\dagger} b&=-\Phi(\mathcal{N})\nonumber,\\
b b^{\dagger}&=-\Phi(\mathcal{N}+1),\label{eq:4.5}
\end{align}
where $\mathcal{N}$ is the number operator and $\Phi(x)$ is a well behaved real function known as the structure function satisfying the following conditions
\begin{align}
\Phi(0)=0, \qquad\text{and} \qquad  \Phi(x)>0, \qquad \text{for}\qquad x>0\label{eq:4.6}. 
\end{align}
 A Fock-space type description is possible when we impose the existence of an integer p such that $\Phi(p+1)=0$.The deformed oscillator algebra in this instance is a parafermionic oscillator algebra.
The realization of this quadratic algebra Q(3) is of the form
\begin{align}
\text{A}=A(\mathcal{N}),\qquad    \text{B}=b(\mathcal{N})+b^\dagger\rho(\mathcal{N})+\rho(\mathcal{N})b,\label{eq:4.7}
\end{align}
where the functions $A(x)$, $b(x)$ and $\rho(x)$ are to be determined.
By defining
\begin{align}
\Delta A(\mathcal{N})= A(\mathcal{N}+1)-A(\mathcal{N}),\label{eq:4.8}
\end{align}
the generator C is realized as 
\begin{align}
C=[A,B]=b^\dagger\Delta A(\mathcal{N})\rho(\mathcal{N})-\rho(\mathcal{N})\Delta A(\mathcal{N})b.\label{eq:4.9}
\end{align}
Now by using (\ref{eq:4.2}), (\ref{eq:4.7}) we have the following two equations 
\begin{align}
(\Delta A(\mathcal{N}))^2=\gamma(A(\mathcal{N}+1)+A(\mathcal{N}))+\epsilon, \label{eq:4.10}\\
\alpha A(\mathcal{N})^2+2 \gamma A(\mathcal{N})b(\mathcal{N})+\delta A(\mathcal{N})+\epsilon b(\mathcal{N})+\zeta=0.\label{eq:4.11}
\end{align}
The function $A(\mathcal{N})$ is determined from (\ref{eq:4.10}) which depends on the value of the parameter $\gamma$ while the $b(\mathcal{N})$ is uniquely obtained from the (\ref{eq:4.11}) provided that atmost one of the parameters $\gamma$ or $\epsilon$ is not zero.

The value of $A(\mathcal{N})$ and $b(\mathcal{N})$ from equation (\ref{eq:4.10}) and (\ref{eq:4.11}) for $\gamma\neq0$ is 
\begin{align}
A(\mathcal{N})&=  \frac{\gamma}{2}((\mathcal{N}+u)^2-\frac{1}{4}-\frac{\epsilon}{\gamma^2}),\label{eq:4.12}\\
b(\mathcal{N})&=-\frac{\alpha((\mathcal{N}+u)^2-\frac{1}{4})}{4}+\frac{\alpha\epsilon-\delta\gamma}{2\gamma^2 }-\frac{\alpha\epsilon^2-2\epsilon\delta\gamma+4\gamma^2 \zeta}{4\gamma^4} \frac{1}{((\mathcal{N}+u)^2-\frac{1}{4}) }\label{eq:4.13},
\end{align}
where the parameter u is to be determined.
For the case when $\gamma=0$ and $\epsilon \neq 0$ we refer the reader to reference [\onlinecite{daskaloyannis2001quadratic}] for the corresponding formulas.

Using (\ref{eq:4.3}), (\ref{eq:4.4}) and the derived value of $A(\mathcal{N})$ and $b(\mathcal{N})$ we arrive at the following two equations
\begin{gather}
2\Phi(\mathcal{N}+1)(\Delta A(\mathcal{N})+\frac{\gamma}{2})\rho(\mathcal{N})-2\Phi(\mathcal{N})(\Delta A(\mathcal{N}-1)
-\frac{\gamma}{2})\rho(\mathcal{N}-1)\nonumber\\=aA^2 (\mathcal{N})
-\gamma b^2 (\mathcal{N})-2\alpha A(\mathcal{N})b(\mathcal{N})+dA(\mathcal{N})-\delta b(\mathcal{N})+z,\label{eq:4.14}
\intertext{and,}
\begin{split}\label{eq:4.15}
K={}&\Phi(\mathcal{N}+1)(\gamma^2-\epsilon-2\gamma A(\mathcal{N})\\
&-\Delta A^2 (\mathcal{N}))\rho(\mathcal{N})+\Phi(\mathcal{N})(\gamma^2-\epsilon-2\gamma A(\mathcal{N})\\
&-\Delta A^2 (\mathcal{N}-1))\rho(\mathcal{N}-1)-2\alpha A^2 (\mathcal{N})b(\mathcal{N})+(\gamma^2-\epsilon-2\gamma A(\mathcal{N})) b^2 (\mathcal{N})\\
&+2(\alpha\gamma-\delta)A(\mathcal{N})b(\mathcal{N})+(\gamma\delta-2\zeta)b(\mathcal{N})+\frac{2}{3} aA^3 (\mathcal{N})\\
&+(d+\frac{1}{3} a\gamma+\alpha^2 ) A^2 (\mathcal{N})+(\frac{1}{3} a\epsilon+\alpha\delta+2z)A(\mathcal{N}).
\end{split}
\end{gather}
 The two equations (\ref{eq:4.14}), (\ref{eq:4.15}) are both linear functions of the expression $\Phi(\mathcal{N})$ and $\Phi(\mathcal{N}+1)$ from which $\Phi(\mathcal{N})$ can be found and $\rho(\mathcal{N})$ can be arbitrarily determined for which the corresponding structure function $\Phi(\mathcal{N})$is a polynomial.
Hence the solution of the function $\Phi(N)$ which depends on the two parameter u and K is given by two different formulas depending on the value of the parameter $\gamma$.
\begin{gather}
\intertext{For $\gamma\neq0$ we have}
\rho(\mathcal{N})=\frac{1}{3.2^{12}.\gamma^8 (\mathcal{N}+u)(\mathcal{N}+u+1)(1+2(\mathcal{N}+u))^2 },\label{eq:4.16}\\
\intertext{and,}
\begin{split}\label{eq:4.17}
\Phi(\mathcal{N})={}&-3072\gamma^6K(-1+2(\mathcal{N}+u))^2\\
&-48\gamma^6(\alpha^2\epsilon-\alpha\delta\gamma+a\epsilon\gamma-d\gamma^2)(-3+2(\mathcal{N}+u))(-1+2(\mathcal{N}+u))^4(1+2(\mathcal{N}+u))\\
&+\gamma^8(-3+2(\mathcal{N}+u))^2(-1+2(\mathcal{N}+u))^4(1+2(\mathcal{N}+u))^2\\
&+768(\alpha\epsilon^2-2\delta\epsilon\gamma+4\gamma^2\zeta)^2+32\gamma^4(-1+2(\mathcal{N}+u))^2(-1+12(\mathcal{N}+u)\\
&+12(\mathcal{N}+u)^2)(3\alpha^2\epsilon^2-6\alpha\delta\epsilon\gamma+2a\epsilon^2\gamma+2\delta^2\gamma^2-4d\epsilon\gamma^2+8\gamma^3z+4\alpha\gamma^2\zeta)\\
&-256\gamma^2(-1+2(\mathcal{N}+u))^2(3\alpha^2\epsilon^3-9\alpha\delta\epsilon^2\gamma+a\epsilon^3\gamma+6\delta^2\epsilon\gamma^2-3d\epsilon^2\gamma^2\\
&+2\delta^2\gamma^4+2d\epsilon\gamma^4+12\epsilon\gamma^3z-4\gamma^5z+12\alpha\epsilon\gamma^2\zeta-12\delta\gamma^3\zeta+4\alpha\gamma^4\zeta).
\end{split}
\end{gather}
The corresponding formulas for $\gamma = 0$ and $\epsilon \neq 0$ are given in reference  [\onlinecite{daskaloyannis2001quadratic}]. 
The Casimir operator $K$ can be expressed in terms of the Hamiltonian H alone. An energy dependent Fock space exists if
\begin{align}
\Phi(p + 1, u, E) = 0,\qquad \Phi(0, u, E) = 0,\qquad \Phi(x) > 0,\label{eq:4.18}
\end{align}
for any positive integer $x=1,2...p$.
The Fock space is then defined by
\begin{align}
H|E, n >= E|E, n >,\qquad \mathcal{N}|E, n >= n|E, n >,\qquad b|E, 0 >= 0,\nonumber\\
b^{\dagger}|n>=\sqrt{\Phi(n+1, E)}|E, n+1>,\qquad b|n>=\sqrt{\Phi(n, E)}|E, n-1>.
\end{align}
The system given by (\ref{eq:4.18}) represents a finite-dimensional unitary representation of the dimension p+1. The solutions of the parameter u and energy E corresponding to the representation of the parafermionic algebra of dimension $p+1$ are determined by (\ref{eq:4.18}).

\section{Quadratic associative algebra and the energy eigenvalues for the 2D quantum harmonic oscillator}\label{Quadratic associative algebra and the energy eigenvalues for the 2D quantum harmonic oscillator}
The 2D quantum Hamiltonian for the harmonic oscillator on the sphere and the hyperbolic plane is given by
\begin{align}
H = \frac{1}{2}\left[\hat{P_x^2}+\hat{P_y^2}+\kappa\hat{J^2}\right] + \frac{\omega^2}{2}\left(\frac{r^2}{1-\kappa r^2}\right),\label{eq:5.1}
\end{align}
where $r^2=x^2+y^2$. Here, $\hat{P_x}$, $\hat{P_y}$, $\hat{J}$ are the quantum version of the Noether momenta arising from the three symmetries of the kinetic energy term. They are given by
\begin{align}
\hat{P_x} = -\iota\hslash\sqrt{1-\kappa r^2}\frac{\partial}{\partial x},\label{eq:5.2}\\  \hat{P_y} = -\iota\hslash\sqrt{1-\kappa r^2}\frac{\partial}{\partial y},\label{eq:5.3}\\
\hat{J} = -\iota\hslash\left(x\frac{\partial}{\partial y}-y\frac{\partial}{\partial x}\right).\label{eq:5.4}
\end{align}
It is possible to decompose the Hamiltonian into the following form
\begin{align}
\hat{H} =& \hat{H_1} + \hat{H_2} +\kappa\hat{J_{12}}^2, \label{eq:5.5}
\intertext{where}
\hat{H_1} =& \frac{1}{2}\left(\hat{P_x}^2+\omega^2\frac{x^2}{1-\kappa r^2}\right)\label{eq:5.6},\\
\hat{H_2} =& \frac{1}{2}\left(\hat{P_y}^2+\omega^2\frac{y^2}{1-\kappa r^2}\right),\label{eq:5.7}\\
\hat{J_{12}}^2 =& -\frac{\hslash^2}{2}\left(x\frac{\partial}{\partial y}-y\frac{\partial}{\partial x}\right)\left(x\frac{\partial}{\partial y}-y\frac{\partial}{\partial x}\right)\label{eq:5.8}.
\end{align}
It is worth noting that the Hamiltonian \^{H} commutes with each of these three terms for any value of $\kappa$ 
\begin{align}
[\hat{H},\hat{H_1}]=[\hat{H},\hat{H_2}]=[\hat{H},\hat{J_{12}}]=0.\label{eq:5.9}    
\end{align}
The vanishing commutators symbolize that the $\kappa$ dependent Hamiltonian is a quantum superintegrable system.
Let us take two independent motion integral operators
\begin{align}
A &= 2\hat{H_1} = \left(\hat{P_x}^2+\omega^2\frac{x^2}{1-\kappa r^2}\right)\label{eq:5.10},\\
B&=2\hat{J_{12}}^2=-\hslash^2\left(x\frac{\partial}{\partial y}-y\frac{\partial}{\partial x}\right)\left(x\frac{\partial}{\partial y}-y\frac{\partial}{\partial x}\right)\label{eq:5.11}.
\end{align}
We can now construct $C$ from $A$ and $B$. Let us consider the quantum superintegrable system under consideration for which we can verify the quadratic associative algebra 
\begin{align}
[A,B]=&C\label{eq:5.12},\\
[A,C]=&8\hslash^2A^2+8\hslash^2\kappa\{A,B\}
-8\hslash^2(2H+\hslash^2\kappa)A+16\hslash^2(\omega^2-\hslash^2\kappa^2)B
\nonumber\\
&+(16\hslash^4\kappa H+8\hslash^4\omega^2),\label{eq:5.13}\\ 
[B,C]=&-8\hslash^2\kappa B^2-8\hslash^2\{A,B\}+16\hslash^4A+8\hslash^2(2H+\hslash^2\kappa)B-16\hslash^4H\label{eq:5.14}.
\end{align}
The Casimir operator for this algebra as determined from (\ref{eq:4.4}) can be expressed in terms of Hamiltonian alone as 
\begin{align}
K = -48\hslash^4H^2-32\hslash^6\omega^2-96\hslash^6\kappa H \label{eq:5.15}.   
\end{align}
The structure function of the associated deformed oscillator can be written as the product of four terms
\begin{align}
    \Phi(x, u, E) = 3221225472 \hbar^{12} \Phi_1(x, u) \Phi_2(x, u, E) \Phi_3(x, u) \Phi_4(x, u, E), \label{eq:5.16}
\end{align}
where the factor functions are
\begin{align}
      \Phi_{1}(x, u) &= -\omega^2 - 2 \kappa^2 \hbar^2(x + u)  + 4\kappa^2 \hbar^2(x + u)^2,  \label{eq:5.17}\\
    \Phi_{2}(x, u, E) &= -2 E \kappa -\omega^2 - 2 \kappa^2 \hbar^2 (x + u)  + 4\kappa^2 \hbar^2 (x + u)^2,  \label{eq:5.18}\\
    \Phi_{3}(x, u) &= -\omega^2 + 2 \kappa^2 \hbar^2 - 6\kappa^2 \hbar^2 (x + u)  + 4 \kappa^2 \hbar^2(x + u)^2, \label{eq:5.19}\\
     \Phi_{4}(x, u, E) &= -2E \kappa -\omega^2 + 2 \kappa^2 \hbar^2 - 
 6 \kappa^2 \hbar^2 (x + u)  + 4\kappa^2 \hbar^2 (x + u)^2 .\label{eq:5.20}
\end{align}
The existence of a finite-dimensional unitary representation of the quadratic algebra of dimension $p+1$ is equivalent to the restrictions (\ref{eq:4.18}) on the annihilation of the structure function for $x=0$ and $x=p+1$, combined with its positivity for $x=1,2...p$. Solving this system of two equations for the two unknowns $E$ and $u$, one obtains the energy eigenvalues. 
The four energy cases given below arise by using the formulas $\Phi_1(0, u)=0$ or $\Phi_3(0,u)=0$ for determining the unknown parameter $u$ and $\Phi_2(p+1,u,E) = 0$ or $\Phi_4(p+1,u,E) = 0$ for determining the energy eigenvalue E. In this way, we obtain the values of the parameter $u$
\begin{align}
 u_{1} &= \frac{1}{4} \pm \frac{1}{4}\sqrt{1 + \frac{4 \omega^2}{\kappa^2 \hbar^2}},\label{eq:5.21}\\
 u_{3} &= u_{1} + \frac{1}{2}\label{eq:5.22}.
\end{align}
Note that the appropriate root with positive sign (negative sign) is applicable when $4 u_{1}$ is greater (smaller) than one. Similarly, when $4 u_{3}$ is greater(smaller) than three then appropriate positive (negative) sign is chosen. The values of the energy eigenvalues are obtained from
\begin{align}
    p+1+u &= \frac{1}{4} \pm \frac{1}{4} \sqrt{\frac{8 E}{\hbar^{2}\kappa}+1+\frac{4 \omega^2}{\hbar^{2}\kappa^{2}}},\label{eq:5.23}\\
    p+1+u &= \frac{3}{4} \pm \frac{1}{4} \sqrt{\frac{8 E}{\hbar^{2}\kappa}+1+\frac{4 \omega^2}{\hbar^{2}\kappa^{2}}}.\label{eq:5.24}
\end{align}
A straightforward tedious calculation shows that
\begin{align}
      E_{n} &= \frac{\hbar^2 \kappa}{2} \left(n+1\right) \left( n + 4 u_{1} \right).\label{eq:5.25}
\end{align}
The relation between $n$ and $p$ is given in Table \ref{table:1} in all the four cases considered above along with the expressions for the factor functions $\Phi_i(i=1,2,3,4)$. As a parenthetical remark, the relation between $n$ and $p$ can also be understood if we substitute the formula (\ref{eq:5.25}) for $E_{n}$ in (\ref{eq:5.23})  and (\ref{eq:5.24}). The quantity under the radical sign becomes a perfect square
\begin{align}
    \sqrt{\frac{8 E_{n}}{\hbar^{2}\kappa}+1+\frac{4 \omega^2}{\hbar^{2}\kappa^{2}}} =  (2 n + 4 u_{1} + 1)^2.
    \label{eq}
\end{align}
Further discussion of the structure function which provides the energy spectra is given in the following section.
\begin{table}[H]
\setlength{\tabcolsep}{4pt}
\centering
\begin{tabular}{||c c c c c ||} 
 \hline
 \makecell{Case \\ $n$} & $\Phi_{1}$ & $\Phi_{2}$ & $\Phi_{3}$ & $\Phi_{4}$ \\ [2ex] 
 \hline
\makecell{Case 1 \\$2p+1$} &  $2x\hbar^{2} \kappa^{2} $ & $(2x-1-n) \hbar^{2} \kappa^{2}$  &  $(2x-1) \hbar^{2} \kappa^{2}$ & $(2x-2-n) \hbar^{2} \kappa^{2}$\\
    & $\times$ $(2x + 4u_{1} -1)$ & $\times$ $(n+2x+4u_{1}) $  & $\times$  $(2x-2+4u_{1})$ & $\times$ $(n+2x+4u_{1}-1)$\\[5ex]

\makecell{Case 2 \\$2p$} &  $2x\hbar^{2} \kappa^{2} $ & $(2x-1-n) \hbar^{2} \kappa^{2}$  &  $(2x-1) \hbar^{2} \kappa^{2}$ & $(2x-2-n) \hbar^{2} \kappa^{2}$\\
    & $\times$ $(2x + 4u_{1} -1)$ & $\times$ $(n+2x+4u_{1}) $  & $\times$ $(2x-2+4u_{1})$ & $\times$ $(n+2x+4u_{1}-1)$\\[5ex]
   
\makecell{Case 3 \\$2p+2$} &  $(2x+1)\hbar^{2} \kappa^{2} $ & $(2x-n) \hbar^{2} \kappa^{2}$  &  $(2x) \hbar^{2} \kappa^{2}$ & $(2x-1-n) \hbar^{2} \kappa^{2}$\\
 & $\times$ $(2x + 4u_{1})$ & $\times$ $(n+2x+4u_{1} + 1) $  & $\times$  $(2x+4u_{1}-1)$ & $\times$ $(n+2x+4u_{1})$\\[5ex]

\makecell{Case 4 \\$2p+1$} &  $(2x+1)\hbar^{2} \kappa^{2} $ & $(2x-n) \hbar^{2} \kappa^{2}$  &  $(2x) \hbar^{2} \kappa^{2}$ & $(2x-1-n)  \hbar^{2} \kappa^{2}$\\
  & $\times$ $(2x + 4u_{1})$ & $\times$ $(n+2x+4u_{1} + 1) $  & $\times$  $(2x+4u_{1}-1)$ & $\times$ $(n+2x+4u_{1})$\\

 \hline
\end{tabular}
\caption{For each of the four energy cases the table enumerates the calculated expressions for the factor functions $\Phi_{i}(i=1,2,3,4)$ in the formula (\ref{eq:5.16}) for the structure function of the deformed oscillator. Note that each $\Phi_{i}$ in case 1 (case 3) is identical to the expression for corresponding $\Phi_{i}$ in case 2 (case 4). Also note that each entry in the third/fourth row is obtained from the corresponding entry in first/ second row by the substitution
$x \rightarrow x+\frac{1}{2}$}
\label{table:1}
\end{table}

\section{Discussion and Conclusions}\label{Discussion and Conclusion}
Following algebraic considerations analogous to those given in reference [\onlinecite{daskaloyannis2001quadratic}], we have obtained the energy eigenvalues $E_{n}$ and the value of the parameter $u$ from the requirement that $\Phi_1$ or $\Phi_3$ vanishes for $x=0$ and $\Phi_2$ or $\Phi_4$ vanishes for $x=p+1$. Our results are summarized in Table \ref{table:1}.

The energy eigenvalues are found to be given in terms of the value $u_{1}$ of the parameter $u$ according to the formula (\ref{eq:5.25})
\begin{equation*}
    E_{n} = \frac{\hbar^2 \kappa}{2}(n+1)( n+4 u_{1} ).
\end{equation*}
It may be recalled that $u_{1}$ is related to $\omega$ according to the formula (\ref{eq:5.21}) as
\begin{equation}
    u_{1} = \frac{1}{4} \pm \frac{1}{4}\sqrt{1 + \frac{4 \omega^2}{\hbar^2\kappa^2}}\nonumber .
\end{equation}
The positive (negative) sign is applicable when $u_{1}$ is greater (smaller) than $1/4$. Now $u_{1}$ is the root of the quadratic equation
\begin{equation}
    \omega^2 = 2 \kappa \hbar u_{1}\left(2 \kappa \hbar u_{1} - \kappa\hbar\right),\label{eq:6.1}
\end{equation}
which results from the requirement that $\Phi_{1}$ given by (\ref{eq:5.17}) vanishes for $x=0$. In order to see that our results agree with the known results from analytic considerations we introduce the parameter
\begin{align}
    \beta = 2 \kappa\hbar u_{1}\label{eq:6.2}.
\end{align}
From (\ref{eq:6.1}) and (\ref{eq:6.2}) we see that
\begin{align}
    \omega^2=\beta(\beta-\kappa \hbar).\label{eq:6.3}
\end{align}
This is the condition for quantum solubility given in references [\onlinecite{carinena2007quantum, carinena2007quantum2, quesne2015update}] and [\onlinecite{carinena2004one}]. Note that
\begin{align}
\lim_{\kappa \to 0} \omega^{2} = \beta^{2}. \label{eq:6.4}
\end{align}
In other words, $\beta$ is the frequency of the flat space isotropic oscillator. In terms of $\beta$, the energy eigenvalues are given by\cite{carinena2007quantum,carinena2007quantum2,quesne2015update}
\begin{align}
    E_{n} = \frac{\hbar^{2} \kappa}{2}(n+1) \left(n+\frac{2 \beta}{\kappa \hbar}\right).\label{eq:6.5}
\end{align}
We now see that the Euclidean limit of (\ref{eq:6.5}) correctly reproduces the known results for energy eigenvalues i.e. $(n+1)\hbar\beta$.
The expression for the structure function for the deformed oscillator in terms of the parameter $\beta$ is given by
\begin{align}
    \Phi_{n}(x,\kappa,\frac{\beta}{\kappa \hbar}) =& \hbar^{4}\kappa^{4} P(P-1)Q(Q-1)R(R-1)S(S-1), \label{eq:6.6}
\end{align}
where Table \ref{table:2} gives the expressions for $P$, $Q$, $R$ and $S$ in the four cases considered by us.
\begin{table}[H]
\setlength{\tabcolsep}{8pt}
\centering
\begin{tabular}{|| c c c c c||} 
 \hline
 \makecell{Case \\ $n$} & $P$ & $Q$ & $R$ & $S$ \\ [2ex] 
 \hline
\makecell{Case 1 \\ $2p+1$}&  $2x$ & $(2x-1-n)$ &  $(2x+\frac{2 \beta}{\kappa \hbar}-1)$ & $(2x+\frac{2 \beta}{\kappa \hbar}+n)$\\[5ex]

\makecell{Case 2 \\ $2p$} &   $2x$ & $(2x-1-n)$ &  $(2x+\frac{2 \beta}{\kappa \hbar}-1)$ & $(2x+\frac{2 \beta}{\kappa \hbar}+n)$\\[5ex]

\makecell{Case 3 \\ $2p+2$} &  $(2x+1) $ & $(2x-n) $  &  $(2x + \frac{2 \beta}{\kappa \hbar}) $ & $(2x+\frac{2 \beta}{\kappa \hbar}+n+1) $\\[5ex]

\makecell{Case 4 \\ $2p+1$} &  $(2x+1) $ & $(2x-n) $  &  $(2x + \frac{2 \beta}{\kappa \hbar}) $ & $(2x+\frac{2 \beta}{\kappa \hbar}+n+1) $\\

 \hline
\end{tabular}
\caption{For each of the four energy cases the table enumerates the expressions for $P, Q, R$ and $S$ arising in the formula (\ref{eq:6.6}) for the structure function of the deformed oscillator. Note that the formulas given in the first (third) and the second (fourth) row agree column by column.
Also note that each entry in third/fourth row is obtained from the corresponding entry in first/second row by the substitution
$x \rightarrow x+\frac{1}{2}$. 
It is also worth mentioning that the entries of the third column can be obtained from the corresponding entries of the second column by the substitution $x \rightarrow \left(x - \frac{n+1}{2} \right)$. Each entry in the fifth column is obtained from the corresponding entry in the fourth column by the substitution $x \rightarrow \left(x + \frac{n+1}{2} \right)$. Each entry in the fourth column is obtained from the corresponding entry in the second column by the substitution $x \rightarrow x + \left(\frac{\beta}{\kappa \hbar} - \frac{1}{2}\right)$   
}
\label{table:2}
\end{table}
We have arrived at (\ref{eq:6.6}) from (\ref{eq:5.16}) on dividing out by $3221225472 \hbar^{16} \kappa^{4}$.
In terms of $\omega^{2}$, we have from (\ref{eq:6.2}) and (\ref{eq:5.21}) the formula
\begin{equation*}
    \frac{\beta}{\kappa \hbar} = \frac{1}{2} \pm \frac{1}{2}\sqrt{1 + \frac{4 \omega^2}{\hbar^2\kappa^2}}.
\end{equation*}
Here it is important to note that the positive (negative) sign applies when $\frac{\beta}{\kappa \hbar}$ is greater (smaller) than $0.5$.

In the Euclidean limit (\ref{eq:6.6}) reduces case by case to the corresponding structure functions given in reference [\onlinecite{daskaloyannis2001quadratic}] for the flat space isotropic oscillator considered as an illustration of the algebraic calculus for calculating the energy spectra.
Note that for the flat space oscillator\cite{daskaloyannis2001quadratic} the parameter $\gamma$ (see (\ref{eq:4.3})) vanishes whereas in contrast with this $\gamma \neq 0$ for the curved space oscillator under consideration. 
It is worth pointing out that had we considered the other four energy cases calculated by using the formula $\Phi_{1}(p+1,u)=0$ or $\Phi_{3}(p+1,u)=0$ for determining the unknown parameter $u$ and $\Phi_{2}(0,u,E)=0$ or $\Phi_{4}(0,u,E)=0$ for determining the energy eigenvalues, our results would not have been valid in the $\kappa=0$ limit. For example, corresponding to case $1$ we would have found instead
\begin{align*}
    E_{n} = \frac{\hbar^{2} \kappa}{2} (n-1)(n-4 u_{1}).
\end{align*}
From here it is now clear that had we used (\ref{eq:6.2}) to express $u_1$ in terms of $\beta$, the Euclidean limit would have led us to the result $E_{n} = -(n-1) \hbar \beta$. This is clearly unacceptable. This is the reason why we have considered only the spectra obtained from the factor structure functions given in Table \ref{table:1}, which are calculated by requiring that $\Phi_{1}$ or $\Phi_{3}$ is annihilated for $x=0$ and $\Phi_{2}$ or $\Phi_{4}$  is annihilated for $x=p+1$.

\begin{figure}[H] 
   \begin{subfigure}{0.32\textwidth}
       \includegraphics[scale=0.22]{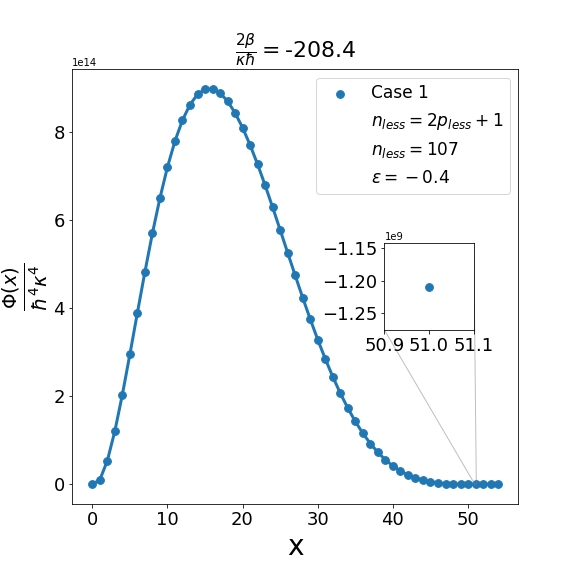}
       \caption{}
   \end{subfigure}
\hfill 
   \begin{subfigure}{0.32\textwidth}
       \includegraphics[scale=0.22]{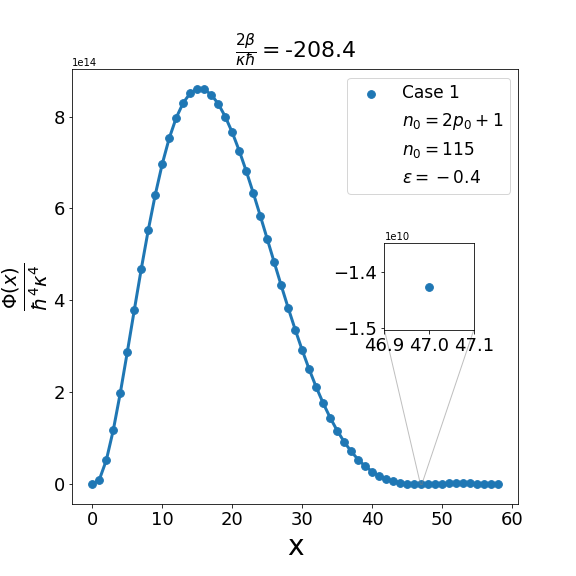}
       \caption{}
   \end{subfigure}
\hfill 
   \begin{subfigure}{0.32\textwidth}
       \includegraphics[scale=0.22]{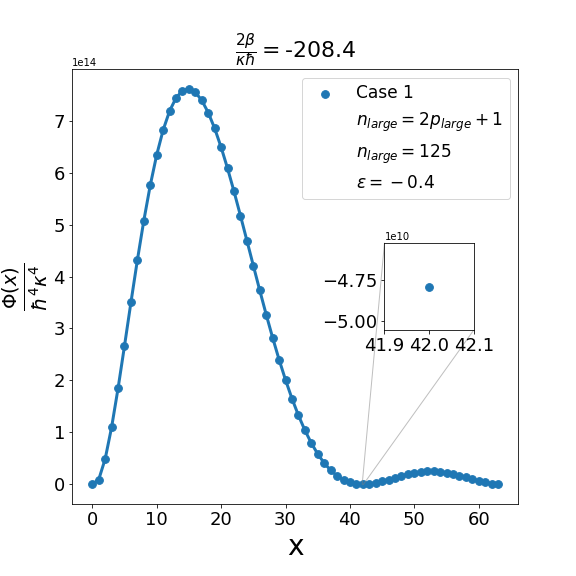}
       \caption{}
   \end{subfigure}
   \caption{The plots shown above exhibit the problem associated with the positivity of the structure function for case 1 given by the formula (\ref{eq:6.6}) with the entries given in first row of Table \ref{table:2}. For $n=n_{0}=115$ the problem arises for $x = x_{0} = 47$. When $n$ is greater (less) than $n_{0}$, the problem appears for $x$ less (greater) than $x_0$. In Fig. \ref{fig:image1} (a), (b) and (c), $S$ evaluates to $1+\epsilon = 0.6$. 
   Moreover, it is apparent from all three plots that $(n_{\text{large}}+2x_{\text{large}}) = (n_{\text{less}}+2x_{\text{less}}) = (n_{0}+2x_{0}) = 209$. Also note that $n_{0}+2x_{0}$ is one more than the integral part of $|\frac{2 \beta}{\kappa \hbar}|$} 
   \label{fig:image1}
\end{figure}

\begin{figure}[H] 
   \begin{subfigure}{0.32\textwidth}
       \includegraphics[scale=0.22]{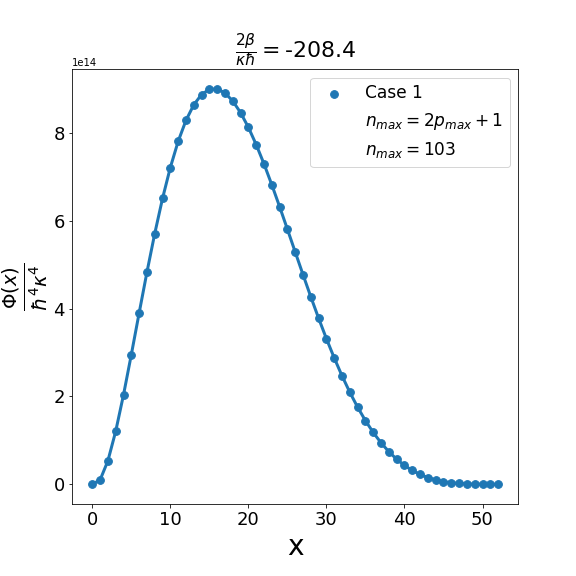}
       \caption{}
   \end{subfigure}
\hfill 
   \begin{subfigure}{0.32\textwidth}
       \includegraphics[scale=0.22]{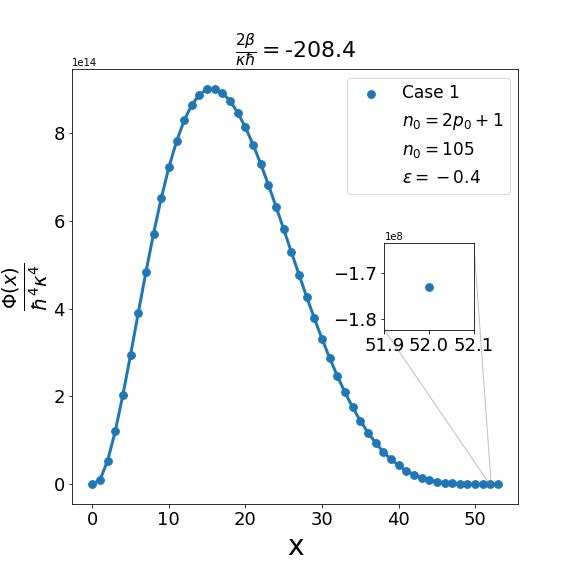}
       \caption{}
   \end{subfigure}
\hfill 
   \begin{subfigure}{0.32\textwidth}
       \includegraphics[scale=0.22]{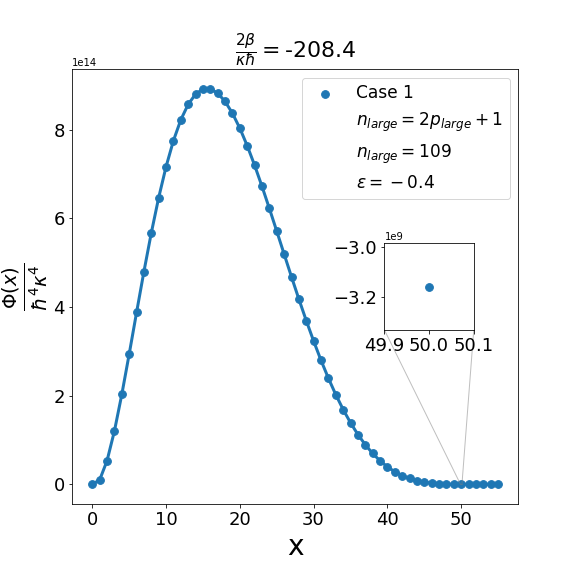}
       \caption{}
   \end{subfigure}
\caption{Plots shown above illustrate the upper bound on $n$ when $\kappa < 0$. Fig. \ref{fig:image2}(b) shows that the value of the structure function for case $1$ for $n=n_{0}=105$ is negative for $x=p_{0} = \frac{n_{0}-1}{2} = 52$. If $n$ exceeds $n_{0}$, the same problem persists as shown in Fig. \ref{fig:image2}(c) for a smaller value of $x$. When $n$ is less than $n_{0}$, the structure function for case 1 is positive for $x=1,2,3...p$, allowing us to impose a bound on $n$, $n_{max} = 103$ as shown in Fig. \ref{fig:image2}(a). This bound agrees with the one given by Quesne\cite{quesne2015update}. In Fig. \ref{fig:image2} (b) and (c), $S$ evaluates to $1+\epsilon = 0.6$. 
Also note that $n_{0}+2x_{0}$ is one more than the integral part of $|\frac{2 \beta}{\kappa \hbar}|$} 
   \label{fig:image2}
\end{figure}

\begin{figure}[H] 
   \begin{subfigure}{0.32\textwidth}
       \includegraphics[scale=0.22]{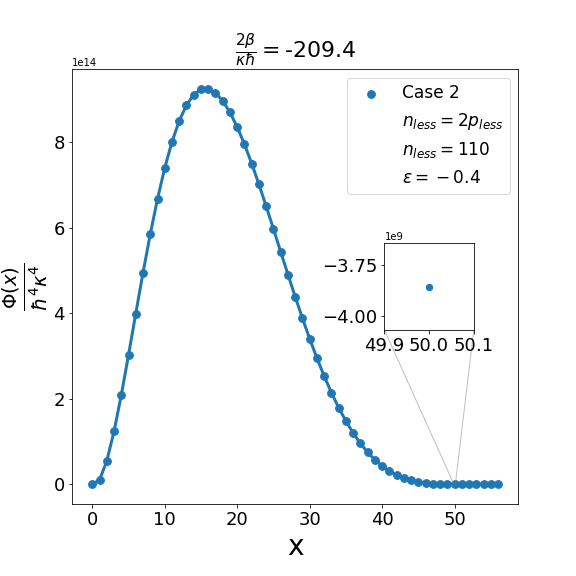}
       \caption{}
   \end{subfigure}
\hfill 
   \begin{subfigure}{0.32\textwidth}
       \includegraphics[scale=0.22]{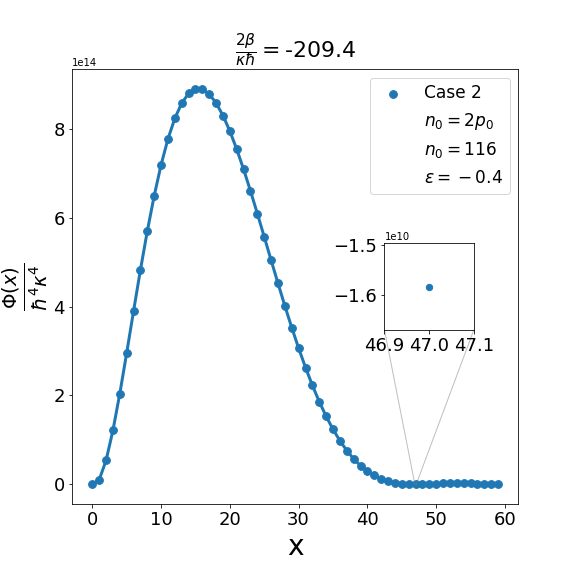}
       \caption{}
   \end{subfigure}
\hfill 
   \begin{subfigure}{0.32\textwidth}
       \includegraphics[scale=0.22]{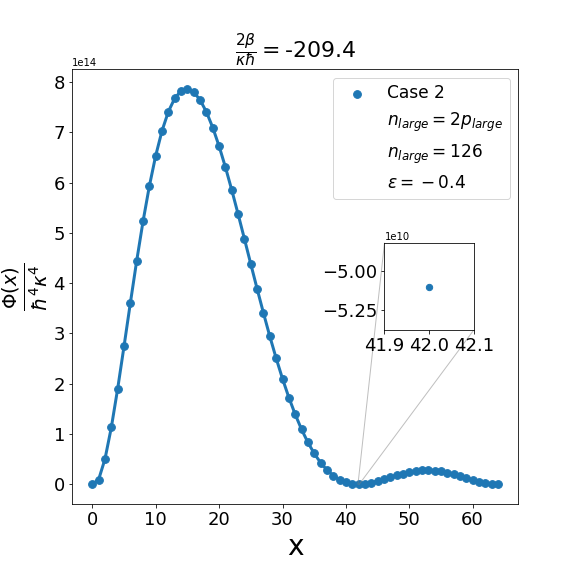}
       \caption{}
   \end{subfigure}
   \caption{The plots shown above exhibit the problem associated with the positivity of the structure function for case 2 given by the formula (\ref{eq:6.6}) with the entries given in second row of Table \ref{table:2}. For $n=n_{0}=116$ the problem arises for $x = x_{0} = 47$. When $n$ is greater (less) than $n_{0}$, the problem appears for $x$ less (greater) than $x_0$. In Fig. \ref{fig:image3} (a), (b) and (c), $S$ evaluates to $1+\epsilon = 0.6$. 
   Moreover, it is apparent from all three plots that $(n_{\text{large}}+2x_{\text{large}}) = (n_{\text{less}}+2x_{\text{less}}) = (n_{0}+2x_{0}) = 210$. Also note that $n_{0}+2x_{0}$ is one more than the integral part of $|\frac{2 \beta}{\kappa \hbar}|$} 
   \label{fig:image3}
\end{figure}

\begin{figure}[H] 
   \begin{subfigure}{0.32\textwidth}
       \includegraphics[scale=0.22]{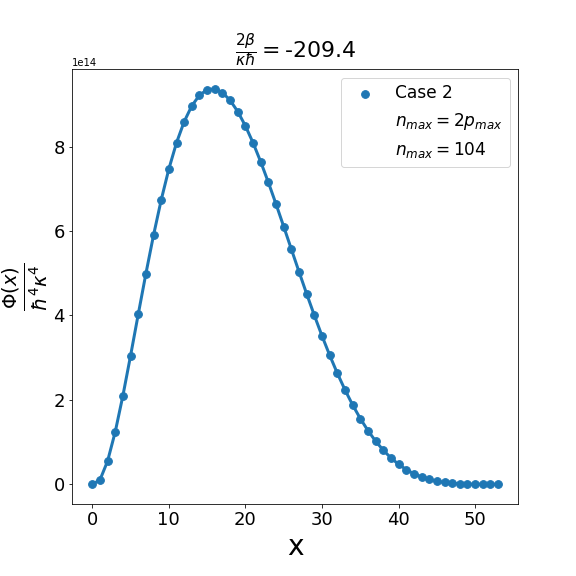}
       \caption{}
   \end{subfigure}
\hfill 
   \begin{subfigure}{0.32\textwidth}
       \includegraphics[scale=0.22]{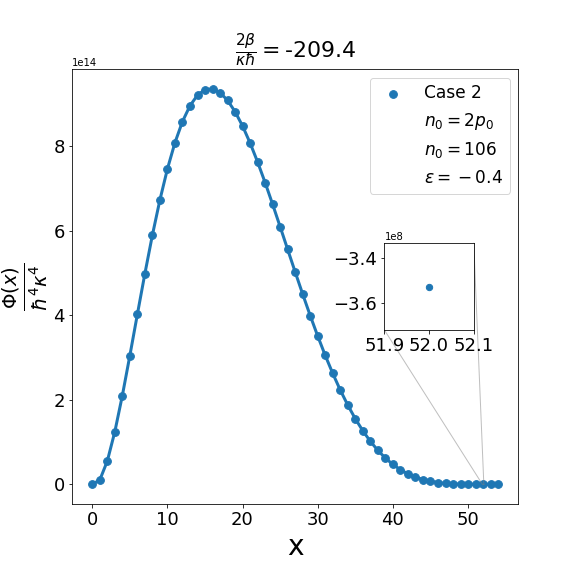}
       \caption{}
   \end{subfigure}
\hfill 
   \begin{subfigure}{0.32\textwidth}
       \includegraphics[scale=0.22]{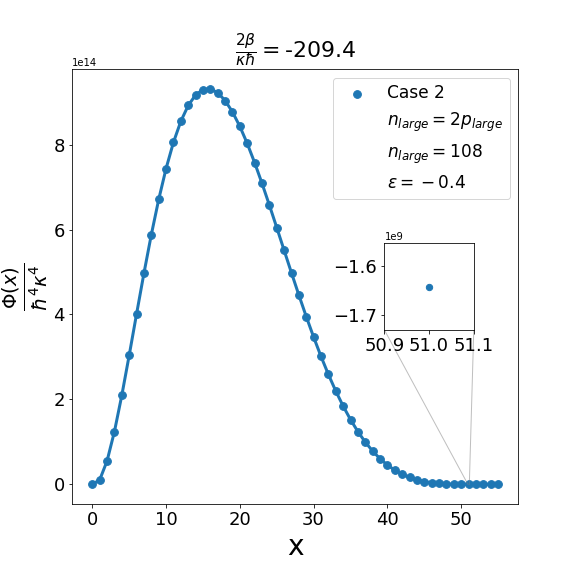}
       \caption{}
   \end{subfigure}
   \caption{Plots shown above illustrate the upper bound on $n$ when $\kappa < 0$. Fig. \ref{fig:image4}(b) shows that the value of the structure function for case $2$ for $n=n_{0}=106$ is negative for $x=p_{0}-1 = \frac{n_{0}}{2} - 1 = 52$. If $n$ exceeds $n_{0}$, the same problem persists as shown in Fig. \ref{fig:image4}(c) for a smaller value of $x$. When $n$ is less than $n_{0}$, the structure function for case 2 is positive for $x=1,2,3...p$, allowing us to impose a bound on $n$, $n_{max} = 104$ as shown in Fig. \ref{fig:image4}(a). This bound agrees with the one given by Quesne\cite{quesne2015update}. In Fig. \ref{fig:image4} (b) and (c), $S$ evaluates to $1+\epsilon = 0.6$. 
Also note that $n_{0}+2x_{0}$ is one more than the integral part of $|\frac{2 \beta}{\kappa \hbar}|$} 
   \label{fig:image4}
\end{figure}

\begin{figure}[H] 
   \begin{subfigure}{0.32\textwidth}
       \includegraphics[scale=0.22]{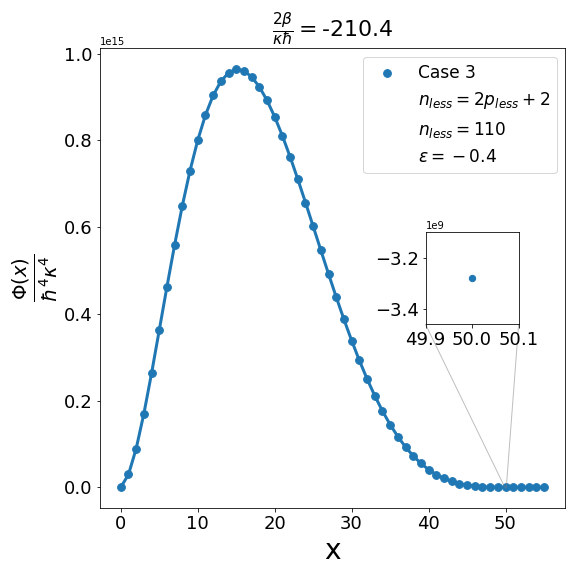}
       \caption{}
   \end{subfigure}
\hfill 
   \begin{subfigure}{0.32\textwidth}
       \includegraphics[scale=0.22]{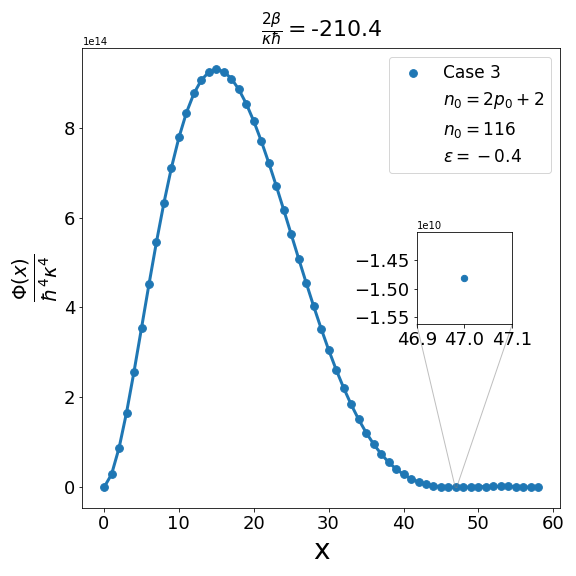}
       \caption{}
   \end{subfigure}
\hfill 
   \begin{subfigure}{0.32\textwidth}
       \includegraphics[scale=0.22]{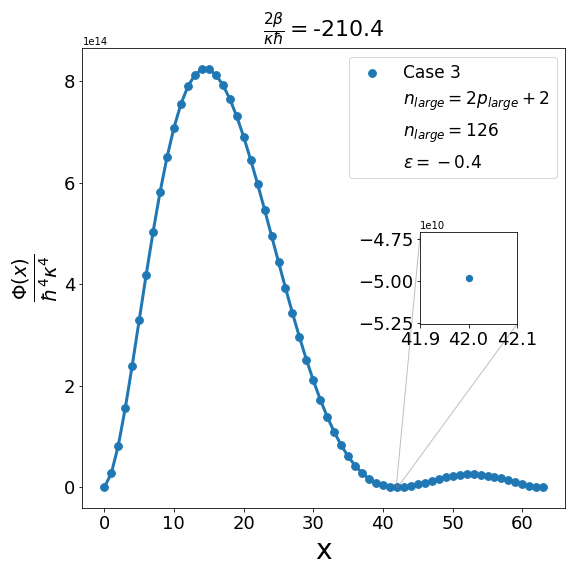}
       \caption{}
   \end{subfigure}
   \caption{The plots shown above exhibit the problem associated with the positivity of the structure function for case 3 given by the formula (\ref{eq:6.6}) with the entries given in third row of Table \ref{table:2}. For $n=n_{0}=116$ the problem arises for $x = x_{0} = 47$. When $n$ is greater (less) than $n_{0}$, the problem appears for $x$ less (greater) than $x_0$. In Fig. \ref{fig:image5} (a), (b) and (c), $S$ evaluates to $1+\epsilon = 0.6$. 
   Moreover, it is apparent from all three plots that $(n_{\text{large}}+2x_{\text{large}}) = (n_{\text{less}}+2x_{\text{less}}) = (n_{0}+2x_{0}) = 210$. Also note that $n_{0}+2x_{0}$ is the integral part of $|\frac{2 \beta}{\kappa \hbar}|$} 
   \label{fig:image5}
\end{figure}

\begin{figure}[H] 
   \begin{subfigure}{0.32\textwidth}
       \includegraphics[scale=0.22]{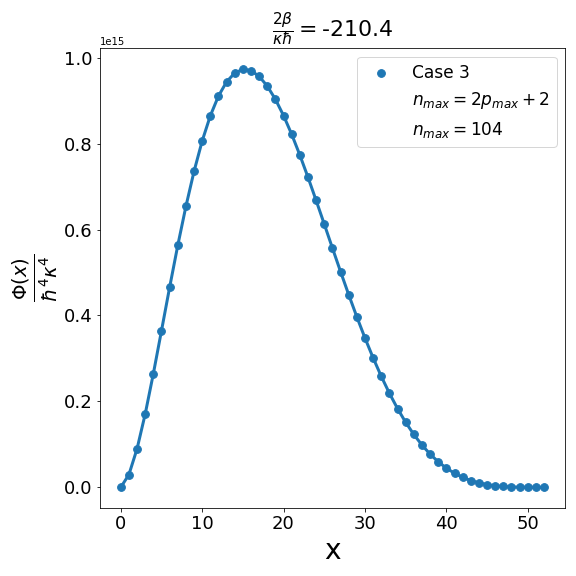}
       \caption{}
   \end{subfigure}
\hfill 
   \begin{subfigure}{0.32\textwidth}
       \includegraphics[scale=0.22]{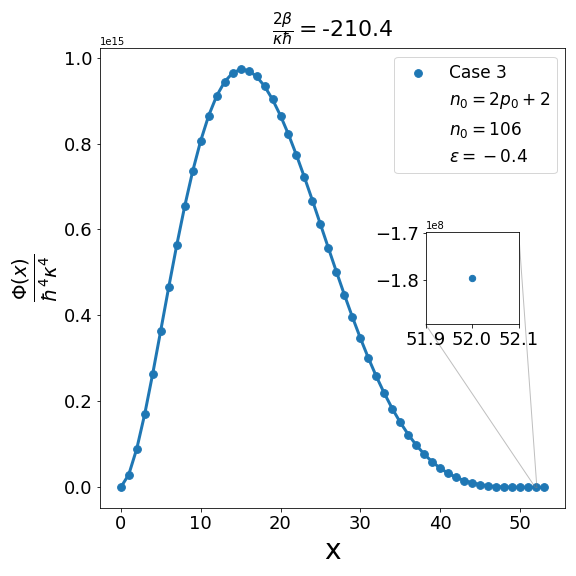}
       \caption{}
   \end{subfigure}
\hfill 
   \begin{subfigure}{0.32\textwidth}
       \includegraphics[scale=0.22]{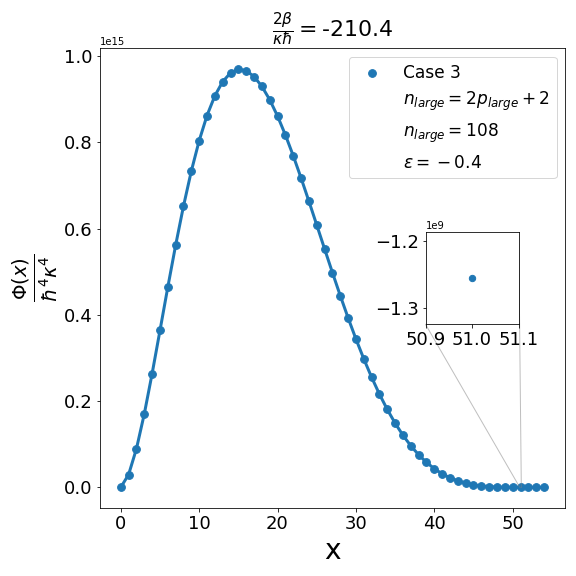}
       \caption{}
   \end{subfigure}
   \caption{Plots shown above illustrate the upper bound on $n$ when $\kappa < 0$. Fig. \ref{fig:image6}(b) shows that the value of the structure function for case $3$ for $n=n_{0}=106$ is negative for $x=p_{0} = \frac{n_{0}-2}{2} = 52$. If $n$ exceeds $n_{0}$, the same problem persists as shown in Fig. \ref{fig:image6}(c) for a smaller value of $x$. When $n$ is less than $n_{0}$, the structure function for case 3 is positive for $x=1,2,3...p$, allowing us to impose a bound on $n$, $n_{max} = 104$ as shown in Fig. \ref{fig:image6}(a). This bound agrees with the one given by Quesne\cite{quesne2015update}. In Fig. \ref{fig:image6} (b) and (c), $S$ evaluates to $1+\epsilon = 0.6$. 
Also note that $n_{0}+2x_{0}$ is the integral part of $|\frac{2 \beta}{\kappa \hbar}|$} 
   \label{fig:image6}
\end{figure}

\begin{figure}[H] 
   \begin{subfigure}{0.32\textwidth}
       \includegraphics[scale=0.22]{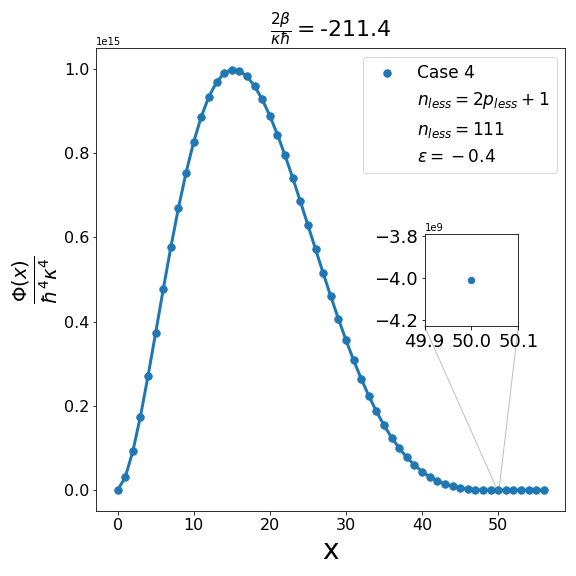}
       \caption{}
   \end{subfigure}
\hfill 
   \begin{subfigure}{0.32\textwidth}
       \includegraphics[scale=0.22]{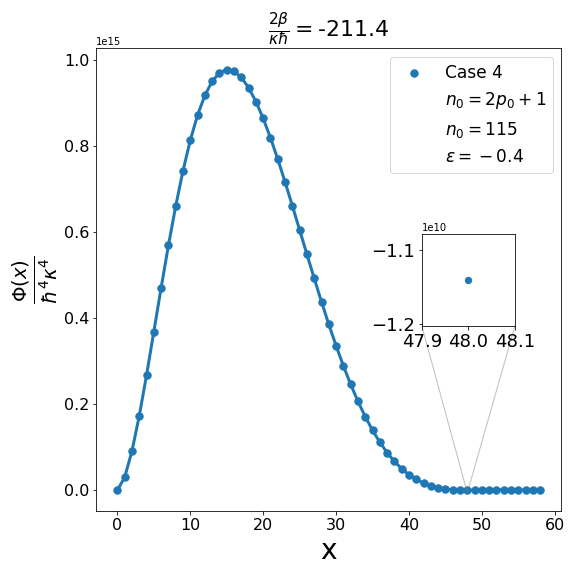}
       \caption{}
   \end{subfigure}
\hfill 
   \begin{subfigure}{0.32\textwidth}
       \includegraphics[scale=0.22]{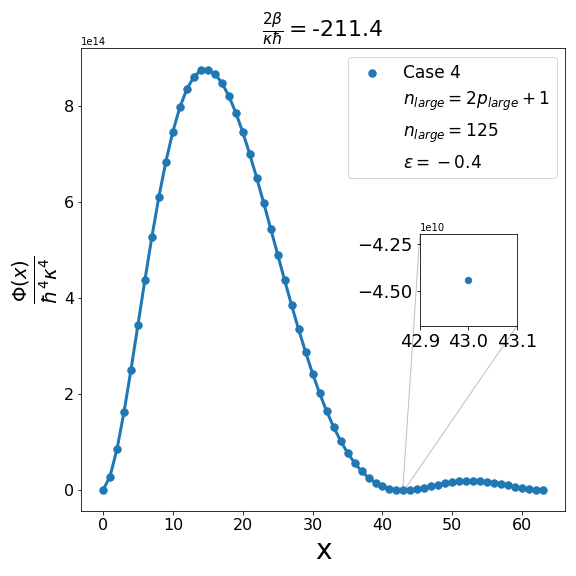}
       \caption{}
   \end{subfigure}
   \caption{The plots shown above exhibit the problem associated with the positivity of the structure function for case 4 given by the formula (\ref{eq:6.6}) with the entries given in fourth row of Table \ref{table:2}. For $n=n_{0}=115$ the problem arises for $x = x_{0} = 48$. When $n$ is greater (less) than $n_{0}$, the problem appears for $x$ less (greater) than $x_0$. In Fig. \ref{fig:image7} (a), (b) and (c), $S$ evaluates to $1+\epsilon = 0.6$. 
   Moreover, it is apparent from all three plots that $(n_{\text{large}}+2x_{\text{large}}) = (n_{\text{less}}+2x_{\text{less}}) = (n_{0}+2x_{0}) = 211$. Also note that $n_{0}+2x_{0}$ is the integral part of $|\frac{2 \beta}{\kappa \hbar}|$} 
   \label{fig:image7}
\end{figure}

\begin{figure}[H] 
   \begin{subfigure}{0.32\textwidth}
       \includegraphics[scale=0.22]{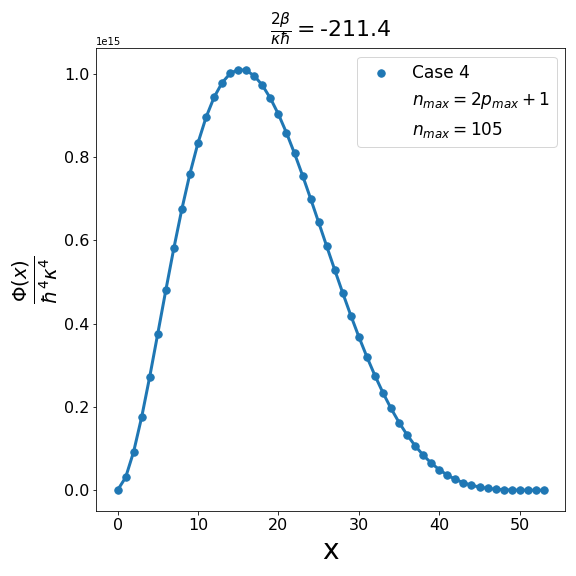}
       \caption{}
   \end{subfigure}
\hfill 
   \begin{subfigure}{0.32\textwidth}
       \includegraphics[scale=0.22]{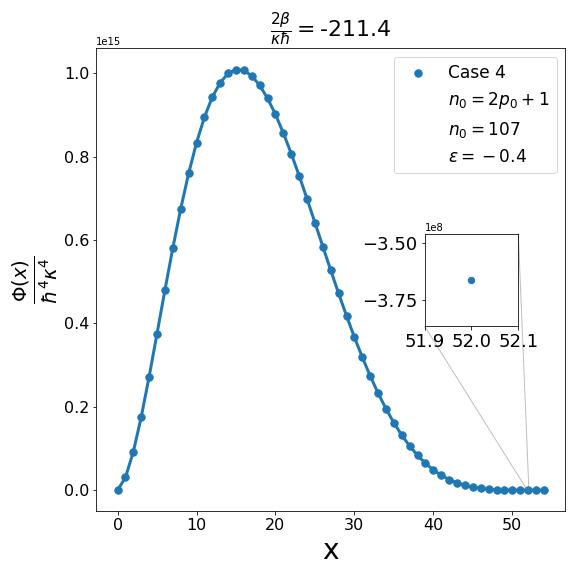}
       \caption{}
   \end{subfigure}
\hfill 
   \begin{subfigure}{0.32\textwidth}
       \includegraphics[scale=0.22]{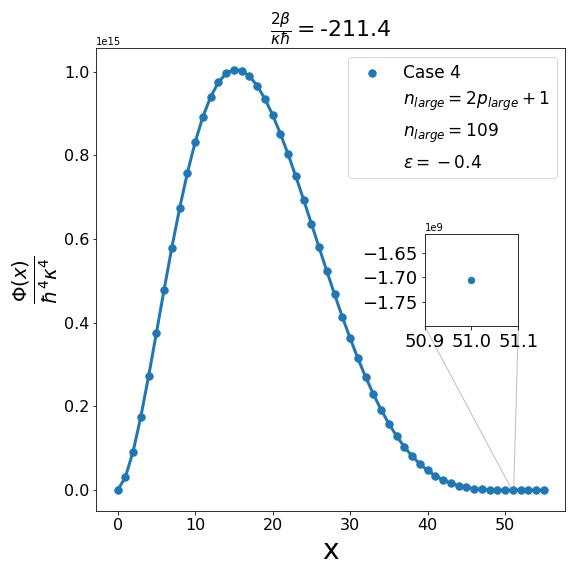}
       \caption{}
   \end{subfigure}
   \caption{Plots shown above illustrate the upper bound on $n$ when $\kappa < 0$. Fig. \ref{fig:image8}(b) shows that the value of the structure function for case $4$ for $n=n_{0}=107$ is negative for $x=p_{0}-1 = \frac{n_{0}-1}{2} - 1 = 52$. If $n$ exceeds $n_{0}$, the same problem persists as shown in Fig. \ref{fig:image8}(c) for a smaller value of $x$. When $n$ is less than $n_{0}$, the structure function for case 4 is positive for $x=1,2,3...p$, allowing us to impose a bound on $n$, $n_{max} = 105$ as shown in Fig. \ref{fig:image8}(a). This bound agrees with the one given by Quesne\cite{quesne2015update}. In Fig. \ref{fig:image8} (b) and (c), $S$ evaluates to $1+\epsilon = 0.6$. 
Also note that $n_{0}+2x_{0}$ is the integral part of $|\frac{2 \beta}{\kappa \hbar}|$} 
   \label{fig:image8}
\end{figure}

Let us now discuss the qualitative difference between the energy spectra on the sphere and on the hyperbolic plane. In the case of the sphere, $\kappa$ is positive and $\frac{\beta}{\kappa \hbar}$ is apparently (much) greater than unity, the positivity of the structure function is not an issue and consequently the energy spectrum is unbounded, i.e. $n$ can be arbitrarily large and therefore there are infinitely many energy eigenvalues. In contrast with this in the other case when $\kappa$ is negative and $\frac{\beta}{\kappa \hbar}$ is a negative quantity of large magnitude, the energy eigenvalues are finite in number because $n$ cannot be arbitrarily large. The actual number of eigenvalues is determined by the positivity of the structure function. Recall that the structure function which is annihilated for $x=0$ and $x=p+1$ must be positive definite for $x=1,2,3...p$. The problem may arise with either $R(R-1)$ or $S(S-1)$ in (\ref{eq:6.6}). If $\frac{-2\beta}{\kappa \hbar} > 2p-1$, we need to worry only about $S(S-1)$. Suppose for a given value of $n=n_{0}$ there exists a positive integral value $x=x_0$ for which $S$ evaluates to $1+\epsilon$, where $\epsilon$ is the signed fractional part of $\frac{2 \beta}{\kappa \hbar}$. The structure function then become negative for $x=x_{0}$. For a value of $n = n_{\text{large}}$ greater than $n_{0}$, the problem reappears for a new value of $x = x_{\text{large}}$ which is less than $x_{0}$. For a value of $n = n_{\text{less}}$ less than $n_{0}$, the problem persists for a new value of $x = x_{\text{less}}$ which is greater than $x_{0}$. As an illustration of the problem which can arise in case 1, we consider typical value $n = n_{0}=115$ (i.e. $p_{0} = 57$) and give the corresponding graph (Fig. \ref{fig:image1}(b)) of the structure function given by formula (\ref{eq:6.6}) with the entries given by the first row of Table \ref{table:2} for $\frac{2 \beta}{\kappa \hbar} = -208.4$, say, i.e. $\epsilon = -0.4$. The structure function becomes negative for $x = x_{0} = 47$. When we decrease the value of $n$ from $n_{0}$ to say $n = n_{\text{less}} = 107$ we find that the structure function becomes negative for $x = x_{\text{less}}=51$. This is shown in Fig. \ref{fig:image1}(a). Similarly, when the value of $n$ is increased to $n = n_{\text{large}} = 125$ it is observed that (see Fig. \ref{fig:image1}(c)) now the problem shifts to $x = x_{large}=42$. Note that as we change the value of $n$, $ (n_{\text{large}} + 2 x_{\text{large}})= (n_{\text{less}} + 2 x_{\text{less}}) = (n_{\text{0}} + 2 x_{\text{0}})$. For the case under consideration $n_{0}+2x_{0}$ evaluates to $209$, which is one more than the integral part of $|\frac{2 \beta}{\kappa \hbar}|$. We may mention in the passing that had we chosen the integral part of $|\frac{2 \beta}{\kappa \hbar}|$ to be odd we would have faced the problem with the positivity of the structure function given by (\ref{eq:6.6}) for half odd integral values of $x$. This is the reason why we opt against this choice.
In order to find the upper bound on $n$ for negative $\kappa$, let us consider the value $n = n_{0}=105$. In this case the plot of the structure function given by formula (\ref{eq:6.6}) with the entries given in the first row of Table \ref{table:2} is shown in Fig. \ref{fig:image2} (b). The graph shows that the problem with positivity arises for $x = x_{0} = p_{0} = \frac{(n_{0}-1)}{2} = 52$. Fig. \ref{fig:image2}(a) shows the plot for $n = 103$. It is observed that there is no problem with the positivity of the structure function for case 1 for integral values of $x=1,2,3...51$. On the other hand for a higher value of $n$ such as $n = n_{\text{large}} = 109$, Fig. \ref{fig:image2}(c) shows that now problem arises for $x = x_{\text{large}} =50$. From the graph it can be seen that $n_{max}$ the maximum allowed value of $n$ is $103$. Note that we have the inequality \cite{quesne2015update}
\begin{align}
    -\frac{\beta}{\kappa \hbar} - \frac{3}{2} < n_{max} < -\frac{\beta}{\kappa \hbar} - \frac{1}{2}. \label{eq:6.7}
\end{align}
We can examine similarly the problem associated with the positivity of the structure functions for cases 2-4 in Table \ref{table:2} as illustrated in Fig. \ref{fig:image3}, Fig. \ref{fig:image5} and Fig. \ref{fig:image7}.
In addition, as shown in Fig. \ref{fig:image4}, Fig. \ref{fig:image6}, and Fig. \ref{fig:image8}, we can determine the maximum allowed value of $n$(i.e. $n_{max}$) for the remaining structure functions (cases 2-4) in Table \ref{table:2}. In each case it is observed that the upper bound on $n$ satisfies the inequality (\ref{eq:6.7}) when $\kappa < 0$. Furthermore contrary to what is true in case 1 and case 2 where $n_{0}+2x_{0}$ is one more than the integral part of $|\frac{2\beta}{\kappa \hbar}|$, $n_{0}+2x_{0}$ in case 3 and case 4 is equal to the integral part $|\frac{2\beta}{\kappa \hbar}|$. 

This article is concerned with the algebraic derivation for spectra of a quantum oscillator on the sphere and on the hyperbolic plane. Starting from the motion integral operators we have constructed the corresponding quadratic algebra satisfied by the integral operators and the Casimir operator of the quadratic algebra. The realisation of the symmetry quadratic algebra in terms of the deformed oscillator enables us to obtain the finite-dimensional unitary representation and the structure functions of the deformed oscillator. The advantage of the algebraic approach is that the structure function is in factorised form which simplifies the calculations significantly. The structure function provides not only the energy spectra algebraically but also a deeper understanding of the number of energy eigenvalues. Our results agree with the known results obtained by classical analytical means.  
\newpage

\newpage
\section{Acknowledgements}
We are indebted to the referee for invaluable comments which helped us significantly in improving the quality of presentation of the revised manuscript. 
\section{DATA AVAILABILITY}
Data sharing is not applicable to this article as no new data were created or analyzed in this study.
\section{References}
\bibliography{aipsamp}
\end{document}